\documentclass[twocolumn]{aastex631}

\usepackage{xcolor}
\usepackage[T1]{fontenc}
\usepackage{microtype}
\usepackage{verbatim}

\hypersetup{
    colorlinks=true,
    linkcolor=blue,
    filecolor=magenta,      
    urlcolor=cyan,
}

\newcommand{\hst}{\textit{HST}}
\newcommand{\HST}{\textit{HST}}

\newcommand{\ha}{\hbox{H$\alpha$}}
\newcommand{\hb}{\hbox{H$\beta$}}

\newcommand{\jwst}{\textit{JWST}}

\newcommand{\grizli}{\textit{Grizli}}

\newcommand{\strt}{\hbox{$S32$}}
\newcommand{\stot}{\hbox{$S3O3$}}
\newcommand{\stt}{\hbox{$S23$}}

\newcommand{\editone}[1]{\textcolor{black}{#1}}
\definecolor{aggiemaroon}{HTML}{500000}
\newcommand{\Sleuth}{\textsc{Sleuth}}
\newcommand{\sleuth}{\textsc{Sleuth}}

\begin{document}

\title{Metal-Poor Star-Forming Clumps in Cosmic Noon Galaxies: Evidence for Gas Inflow and Chemical Dilution Using JWST NIRISS}

\author[0000-0001-8489-2349]{Vicente Estrada-Carpenter}
\affiliation{School of Earth and Space Exploration, Arizona State University, Tempe, AZ 85287, USA}
\affiliation{Beus Center for Cosmic Foundations, Arizona State University, Tempe, AZ 85287, USA}
\affiliation{Institute for Computational Astrophysics and Department of Astronomy \& Physics, Saint Mary's University, 923 Robie Street, Halifax, NS B3H 3C3, Canada}

\author[0000-0002-7712-7857]{Marcin Sawicki}
\affiliation{Institute for Computational Astrophysics and Department of Astronomy \& Physics, Saint Mary's University, 923 Robie Street, Halifax, NS B3H 3C3, Canada}

\author[0000-0002-4542-921X]{Roberto Abraham}
\affiliation{David A. Dunlap Department of Astronomy and Astrophysics, University of Toronto, 50 St. George Street, Toronto, Ontario, M5S 3H4, Canada}

\author[0000-0003-3983-5438]{Yoshihisa Asada}
\affiliation{Waseda Research Institute for Science and Engineering, Faculty of Science and Engineering, Waseda University, 3-4-1 Okubo, Shinjuku, Tokyo 169-8555, Japan}
\affiliation{Institute for Computational Astrophysics and Department of Astronomy \& Physics, Saint Mary's University, 923 Robie Street, Halifax, NS B3H 3C3, Canada}
\affiliation{Department of Astronomy, Kyoto University, Sakyo-ku, Kyoto 606-8502, Japan}

\author[0000-0001-5984-0395]{Maru\v{s}a Brada\v{c}}
\affiliation{Faculty of Mathematics and Physics, Jadranska ulica 19, SI-1000 Ljubljana, Slovenia}
\affiliation{Department of Physics and Astronomy, University of California Davis, 1 Shields Avenue, Davis, CA 95616, USA}

\author[0000-0003-2680-005X]{Gabe Brammer}
\affiliation{Cosmic Dawn Center (DAWN), Denmark}
\affiliation{Niels Bohr Institute, University of Copenhagen, Jagtvej 128, DK-2200 Copenhagen N, Denmark}

\author[0000-0001-8325-1742]{Guillaume Desprez}
\affiliation{Kapteyn Astronomical Institute, University of Groningen, P.O. Box 800, 9700AV Groningen, The Netherlands}
\affiliation{Institute for Computational Astrophysics and Department of Astronomy \& Physics, Saint Mary's University, 923 Robie Street, Halifax, NS B3H 3C3, Canada}

\author[0000-0001-9298-3523]{Kartheik G. Iyer}
\affiliation{Columbia Astrophysics Laboratory, Columbia University, 550 West 120th Street, New York, NY 10027, USA}

\author[0000-0003-3243-9969]{Nicholas S. Martis}
\affiliation{Faculty of Mathematics and Physics, Jadranska ulica 19, SI-1000 Ljubljana, Slovenia}

\author[0000-0002-9330-9108]{Adam Muzzin}
\affiliation{Department of Physics and Astronomy, York University, 4700 Keele St. Toronto, Ontario, M3J 1P3, Canada}

\author{Ga\"el Noirot}
\affiliation{Space Telescope Science Institute, 3700 San Martin Drive, Baltimore, Maryland 21218, USA}

\author[0009-0009-4388-898X]{Gregor Rihtar\v{s}i\v{c}}
\affiliation{Faculty of Mathematics and Physics, Jadranska ulica 19, SI-1000 Ljubljana, Slovenia}

\author[0000-0001-8830-2166]{Ghassan T. E. Sarrouh}
\affiliation{Department of Physics and Astronomy, York University, 4700 Keele St. Toronto, Ontario, M3J 1P3, Canada}

\author[0000-0002-4201-7367]{Chris J. Willott}
\affiliation{National Research Council of Canada, Herzberg Astronomy \& Astrophysics Research Centre, 5071 West Saanich Road, Victoria, BC, V9E 2E7, Canada}

\author[0009-0005-6999-2073]{Jeremy Favaro}
\affiliation{Institute for Computational Astrophysics and Department of Astronomy \& Physics, Saint Mary's University, 923 Robie Street, Halifax, NS B3H 3C3, Canada}

\author[0000-0002-5694-6124]{Vladan Markov}
\affiliation{Faculty of Mathematics and Physics, Jadranska ulica 19, SI-1000 Ljubljana, Slovenia}

\author[0000-0001-8115-5845]{Rosa M. M\'erida}
\affiliation{Institute for Computational Astrophysics and Department of Astronomy \& Physics, Saint Mary's University, 923 Robie Street, Halifax, NS B3H 3C3, Canada}

\author[0009-0009-2307-2350]{Katherine Myers}
\affiliation{Department of Physics and Astronomy, York University, 4700 Keele St. Toronto, Ontario, M3J 1P3, Canada}

\author[0000-0003-0780-9526]{Visal Sok}
\affiliation{Department of Physics and Astronomy, York University, 4700 Keele St. Toronto, Ontario, M3J 1P3, Canada}

\begin{abstract}
The formation and evolution of galaxies are intricately linked to the baryon cycle, which fuels star formation while shaping chemical abundances within galaxies. Investigating the relationship between star formation and metallicity for large samples of galaxies requires expensive IFU surveys or sophisticated tools to analyze grism data. Here we analyze JWST NIRISS slitless grism data using \Sleuth, a tool that forward models and infers spatially resolved physical properties from grism data, including observations from JWST NIRISS/NIRCam and future grism data like that from the Roman Space Telescope. \Sleuth\ enables extraction of high-quality emission line maps from slitless spectra, overcoming contamination and spatially varying stellar populations, which previously limited such studies. Utilizing \Sleuth\ with data from the CAnadian NIRISS Unbiased Cluster Survey (CANUCS), we investigated the relationship between metallicity and star formation in the star-forming clumps of galaxies at 0.6 < z < 1.35. We analyzed a sample of 20 galaxies, extracted high-quality emission line maps with \sleuth, and analyzed, in detail, the spatially resolved properties of star-forming clumps. Using \ha, [SII], and [SIII] emission line maps, we examined the spatially resolved metallicities, ionization, and star formation rates of our sample. Our findings reveal that these star-forming clumps show lower metallicities ($\sim$ 0.1 dex) than their surrounding galactic environments, indicating a metallicity dilution of 20$\%$ within the clumps' gas. Our analysis indicates that these clumps exhibit intensified star formation and reduced metallicity, likely due to the inflow of metal-poor gas. These clumps illustrate the dynamic relationship between star formation and chemical enrichment within galaxies.

\end{abstract}


\section{Introduction}
Star formation is a complex process, and its complexity has been made all the more evident through the high-resolution capabilities of \jwst. The ability to see high redshift galaxies in greater detail has allowed us to see the intricate morphologies of these galaxies, including dense regions of star formation \citep{mowl22, clae23, dunc24,estr24}. These star-forming clumps are evident in the UV and \ha\ emission and can represent a large percentage of a galaxy's overall star formation \citep{wuyt12,mura14,guo15,sok22, satt23, estr24,sok25}.


Studies focusing on low redshift (z $<$ 0.15 )galaxies have examined the metallicities of star-forming clumps \citep{rich14,alme15,kuma17, hwan19, olve24}. Their findings indicated that these clumps have lower metallicities and higher star formation rates when compared to their host galaxies. These finding suggests that inflowing metal-poor gas might be responsible for fueling the star formation seen, and as a consequence, leading to a dilution of metallicity in these regions. Therefore, star-forming clumps are valuable subjects for exploring how chemical enrichment relates to star formation. 

To gain a clearer understanding of how metallicity connects with star formation, it's important to examine higher redshift galaxies. This is because the fraction of clumpy galaxies correlates with the star formation rate density of the universe \citep{mada14}. Studies show that as we look at higher redshifts, the proportion of clumpy galaxies increases, peaking between 1 < z < 2 \citep{satt23}. 

\cite{sok25} looked at clumpy galaxies at z $\sim$ 0.7, and noted that galaxies with star-forming clumps possess overall lower integrated metallicities (0.07 ± 0.02 dex) compared to their non-clumpy counterparts. A key question raised by \cite{sok25} is why clumpy galaxies exhibit overall lower metallicities than their non-clumpy counterparts, and whether this is driven by the formation of the star-forming clumps, similar to what was seen at lower redshift. We can address this question with \jwst/NIRISS Wide-Field Slitless Spectroscopy \citep{will22}, which provides spatially resolved coverage of Cosmic Noon (1 < z < 3) galaxies in the rest-frame optical. Using spatially resolved slitless spectroscopy, we can extract emission line maps that detail the location of emission within the galaxies \citep{simo21,math22,math23,estr24}. Recent innovations in grism analysis have allowed for the ability to extract high-quality, high-resolution emission line maps from the NIRISS grism spectroscopy. These maps allowed for the detailed study of galaxy properties as seen in \citet{estr24}, where we can study the properties of individual star-forming clumps. 

In this work, we use \sleuth, our advanced grism analysis code introduced in \cite{estr24}, to extract high-quality emission line maps from \jwst/NIRISS grism spectra to study the connection between metallicity and star formation in star-forming clumps at Cosmic Noon. In Section \ref{sec:data} we outline the data from the CAnadian NIRISS Unbiased Cluster Survey (CANUCS, \citealp{will22}) used in our study, as well as detail our sample selection. In Section \ref{sec:meth} we discuss the tools and methods used for our spatially resolved analysis, including how we identify clumps and calculate metallicity. In Section \ref{sec:res} we detail our findings on the metallicity of star-forming clumps. Finally, in Section \ref{sec:dis}, we discuss the link between star formation and metallicity as seen in the star-forming clumps.  Throughout this work, we assume a cosmology with $\Omega_{m,0}$ = 0.3, $\Omega_{\lambda,0}$ = 0.7, and $H_{0}$ = 70 km s$^{-1}$.

\section{Data \label{sec:data}}
\subsection{Imaging and Photometry}

\begin{figure*}[ht!]
\plotone{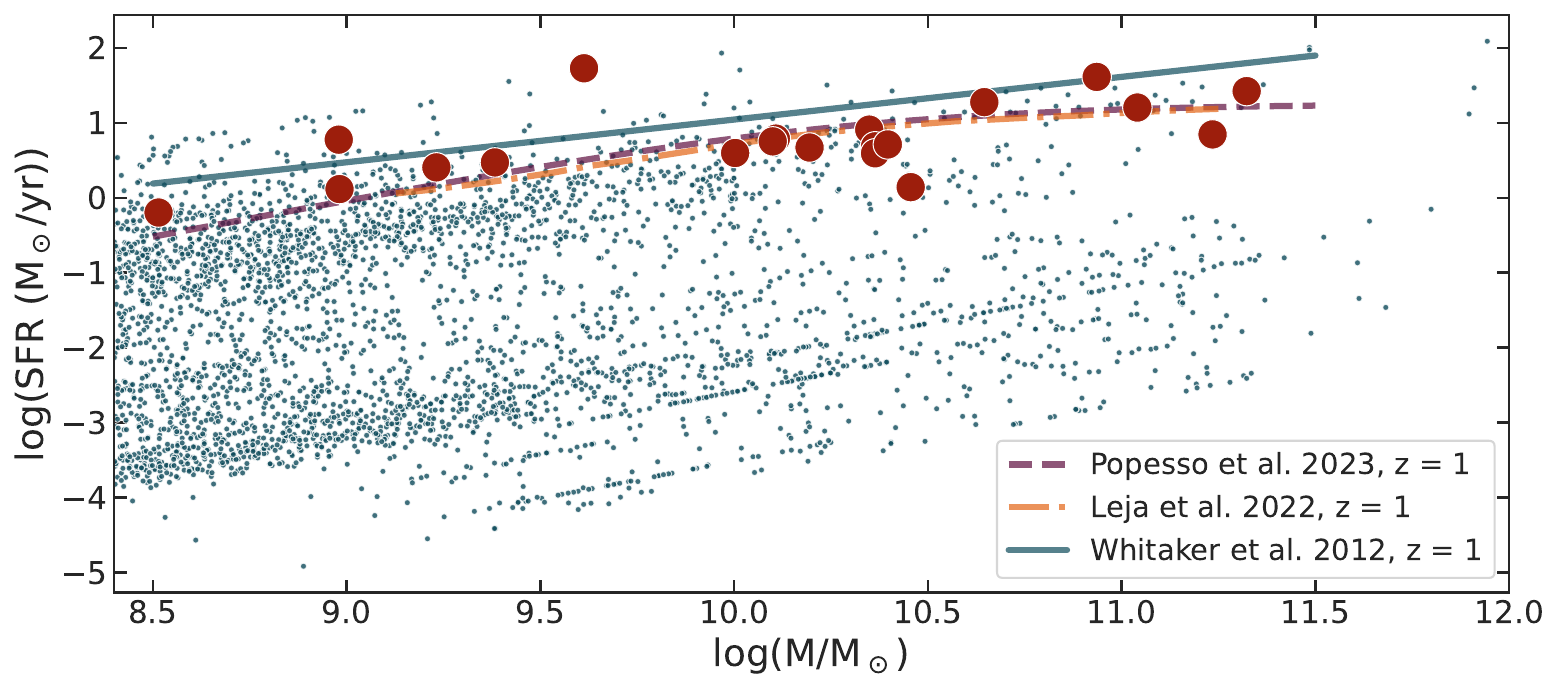}
\caption{The parent sample of galaxies from the MACS J0417, MACS J1149, and MACS J1423 cluster fields, with star formation rates (SFRs) and masses (both corrected for lensing) determined using Dense Basis. Blue dots represent the parent sample, while our selected sample is highlighted as larger red dots. Star-forming main sequence fits at z = 1  from \cite{whit12}, \cite{leja22}, and  \cite{pope23}, are included as a green solid line, orange dash-dot line, and purple dashed line, respectively. Most galaxies analyzed in this study are located close to the star-forming main sequence.  
\label{fig:sample}}
\end{figure*}

We used data from the MACS J0417, MACS J1149, and MACS J1423 cluster fields taken as part of the CANUCS survey \citep{will22}. The CANUCS dataset consists of imaging in \jwst\ NIRCam (F090W, F115W, F150W, F200W, F277W, F356W, F410M, F444W) with exposure times of 6.4 ks each with a signal to noise ratio (SNR) between 5 to 10 for an AB = 29 point source \citep{asad23,stra23, sarr25}). In addition to the \jwst\ data, we incorporated archival \hst\/ACS data (F435W, F606W, F814W). Cluster galaxies were modeled and removed using isophotal models, as outlined in  \cite{mart24}. Point Spread Functions (PSFs) for broadband imaging were empirically determined by median-stacking non-saturated bright stars. NIRCam broadband data consist of PSF-convolved images standardized to the resolution of the \jwst\ NIRCam F444W. Additionally, star formation rates (SFRs) and masses were derived from this dataset and are included in the CANUCS DR1 paper (Sarrouh, Asada et al. in prep). Strong lensing models of the three clusters were derived with Lenstool \citep{jull07} using CANUCS imaging and spectroscopic data. These will be published in Desprez et al in prep. (MACS J0417), Rihtaršič et al. in prep. (MACS J1149), and Desprez et al. in prep (MACS J1423).

\subsection{CANUCS NIRISS Data}
The CANUCS \jwst\ NIRISS grism data were obtained using the GR150R and GR150C grisms and utilized the F115W, F150W, and F200W filters (each having an exposure time of 19.2 ks \citealp{math23}). The data processing was conducted using the grism modeling and analysis software \grizli\ \citep{grizli}, which performs an end-to-end reduction of the data and models contamination, as described in \cite{noir23}. Additional processing was completed to remove the 2nd, 1st, 0th, and -1st order spectra of the cluster galaxies as described in \cite{estr24}. As shown in that paper, by removing the cluster galaxy spectra, we can extract more accurate emission line maps as these foreground cluster galaxies are large and bright, causing their spectra to contaminate many other spectra. 

\subsection{Sample Selection}
The goal of this project is to create metallicity maps for galaxies that exhibit clumpy star formation. To achieve this, we required that our galaxies fall within a redshift range of 0.6 < z$_{grism}$ < 1.35 as this redshift range ensures that the \ha\ and [SIII] emission lines will be captured within the \jwst/NIRISS F115W/F150W/F200W filters. We also require that the galaxies NIRISS grism spectra have H$\alpha$, [SII] $\lambda$ 6730, and [SIII]$\lambda \lambda$ 9069,9532 emission lines detected, as we will use these lines to calculate metallicity.

Additionally, we require that galaxies in our sample have star-forming clumps. We identify these regions using the method outlined in Section \ref{sec:clump}. From the clumpy galaxies identified, we exclude any multiple images of a single galaxy (note that the fields used in this study are lensed by massive clusters) and retain only the highest magnified / least contaminated (in the grism spectra) image, determined by eye. This process yields a final sample of 20 galaxies.

Figure \ref{fig:sample} illustrates the distribution of our selected galaxies in relation to all galaxies (0.5 $<$ z $<$ 1.35 ) in the three CANUCS fields we examine, plotted along the star-forming main sequence \citep{whit12, leja22, pope23} with SFRs and stellar masses (de-lensed) constrained from photometry using Dense Basis (\cite{iyer19}, Sarrouh, Asada et al. in prep). In this figure, the parent sample is represented by blue dots, while the galaxies selected for our study are shown as larger red dots. Most of our target galaxies are positioned close to the main sequence.

\section{Methods \label{sec:meth}}
\begin{figure*}[ht!]
\plotone{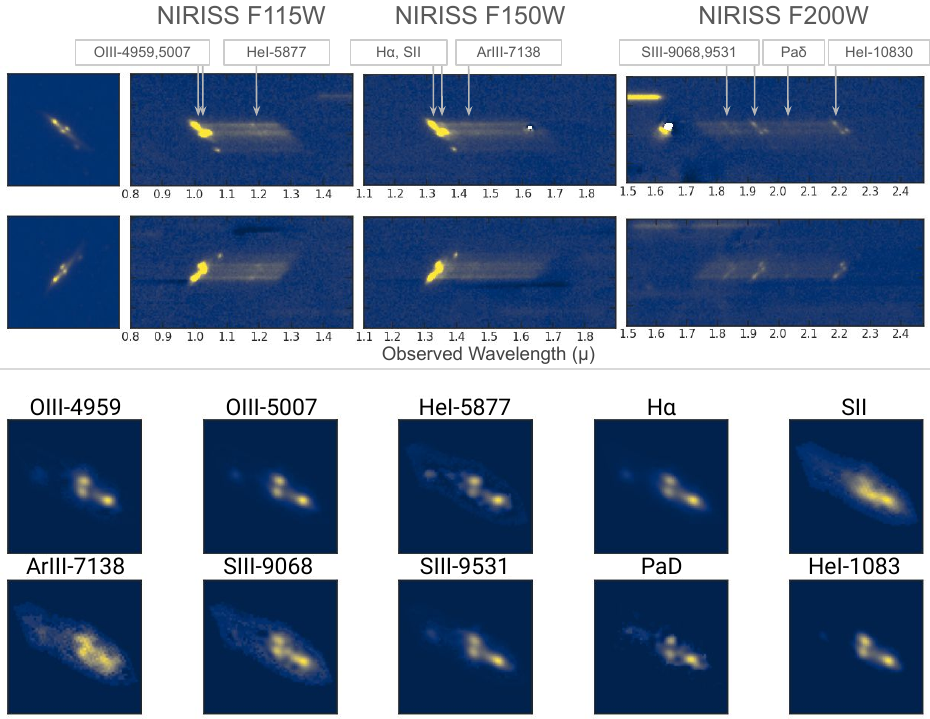}
\caption{Example \jwst/NIRISS grism data and emission line maps from one of our sample galaxies (1100250) at z = 1.043$^{+0.001}_{-0.001}$. The top panel shows the spatially resolved NIRISS grism data in the F115W, F150W, and F200W filters (GR150C and GR150R are shown in the top and bottom rows, respectively). This galaxy exhibits several emission lines, which are highlighted with arrows. The bottom panel shows the extracted emission line maps for this galaxy. While some of these emission lines are contaminated by nearby lines, \Sleuth's spatially resolved fitting is able to extract high-quality maps for all lines. Note that the emission line maps have been binned using \sleuth's segmentation feature and are oriented differently from the NIRISS direct images in the top row. These high-quality maps will allow us to constrain several spatially resolved properties of our galaxies. 
\label{fig:data}}
\end{figure*}

\subsection{Spatially Resolved Broad Band Fits \label{sec:bbfit}}
For each of our galaxies, we derive spatially resolved broadband fits utilizing the PSF matched (to F444W) \HST\ and \jwst\ imaging data from the CANUCS survey. Instead of analyzing the galaxies pixel-by-pixel, we subdivide them into smaller regions based on color and SNR (SNR = 50 in F150W), similar to the method described in \cite{estr24}. As many regions end up being single pixels, this approach allows us to benefit from the high spatial resolution of \jwst. 

We then fit each region using Flexible Stellar Population Synthesis (FSPS) models \citep{conr10} using a combination of MILeS and BaSeL libraries and a Kroupa initial mass function \citep{krou01}, constraining specific star formation rate (sSFR), metallicity, dust, mass, and non-parametric star formation histories from Dense Basis. We fix the redshifts to grism-derived ones as they have smaller uncertainties ($\delta$z $\sim$ 0.01 \citealt{noir23}) than photo-zs.

\begin{figure*}[ht!]
\plotone{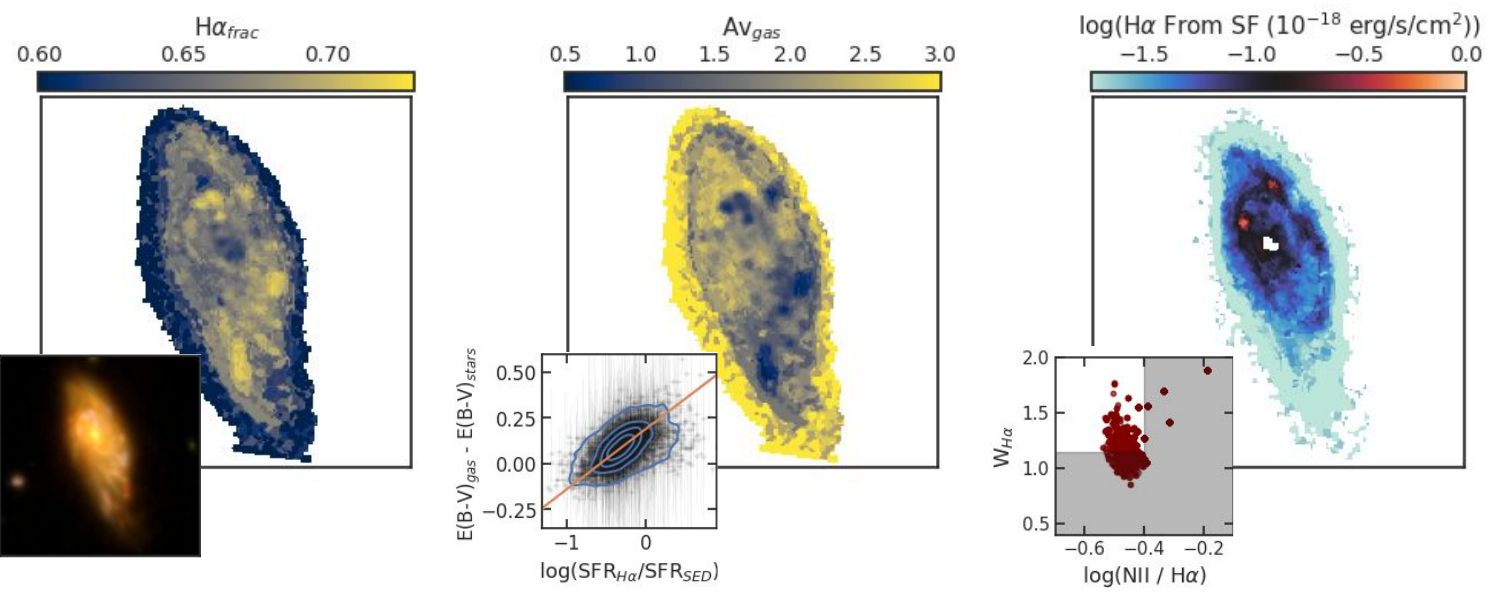}
\caption{\editone{Emission line correction maps. The leftmost panel shows the estimated fraction of \ha\ emission within the blended [NII] + \ha\ maps, derived from dust-corrected B–V color maps using the empirical relation from \citet{sanc12}. The middle panel displays the nebular dust attenuation map, AV$_{gas}$, computed by combining spatially resolved A$V$ maps from the broadband SED fits with an estimate of E(B–V)$_{gas}$ – E(B–V)$_{stars}$, based on the ratio of SFR$_{H\alpha}$ to SFR$_{10 Myr}$ (the relationship shown in the middle inset panel). The rightmost panel indicates regions where \ha\ emission is consistent with star formation, identified using the WHaN diagnostic diagram (shown in the rightmost inset panel).} \label{fig:corr}}
\end{figure*}

\subsection{Emission Line Map Corrections}

\subsubsection{\ha Fraction}
\editone{Before we derive metallicity maps, we apply several corrections to the emission line maps. First, we correct our \ha\ maps. In the NIRISS grism spectra, \ha\ and [NII] lines are always blended. To better estimate metallicity, we must determine the proportion of \ha\ emission in our combined \ha\ + [NII] maps. We achieve this by using dust-corrected B-V maps in the rest frame. As shown in \cite{sanc12} (Fig. B.1), there is a relationship between B-V and log([NII]/\ha).}

\begin{equation}
log([Nii]/H\alpha) = -0.64 + 0.36(B - V)    
\end{equation}

\editone{The leftmost panel of Figure \ref{fig:corr} shows the \ha$_{frac}$ (the fraction of \ha\ emission in the \ha\ + [NII] maps) for galaxy 4101288. Here we see that the \ha$_{frac}$ is lower in the outskirts of the galaxy and in a few small regions in the inner galaxy.}

\subsubsection{Dust Correction}
\editone{To correct for dust attenuation in our emission line maps, we use dust maps derived from spatially resolved broadband SED fitting (see Section \ref{sec:bbfit}). However, since these dust estimates are based on the continuum, they may underestimate the attenuation affecting nebular emission, which can be more heavily obscured \citep{char00}. To estimate E(B–V)$_{gas}$, we calibrate a relationship between E(B–V)$_{gas}$ – E(B–V)$_{stars}$ and the ratio of SFR$_{H\alpha}$ to SFR$_{10 Myr}$. SFR$_{H\alpha}$ is derived from our \ha\ maps using the calibration from \cite{kenn94}, while SFR$_{10 Myr}$ is extracted from the star formation history in each region of the galaxy (see Section \ref{sec:bbfit}), specifically at a lookback time of 10 Myr. This empirical relation is calibrated using measurements from MaNGA \citep{bund15}, as presented in the catalog by \cite{sanc22} and shown in the central inset of Figure \ref{fig:corr}. Since both SFR indicators trace star formation over comparable timescales, discrepancies between them can primarily reflect differences in dust attenuation. The relationship is therefore:}

\begin{equation}
E(B-V)_{gas} - E(B-V)_{stars} = 0.33  \left(\frac{SFR_{H\alpha}}{SFR_{10Myr}}\right)  + 0.19
\end{equation}

\editone{We estimate the effects of dust on our emission line maps using an iterative approach. Initially, we assume a fixed relation between stellar and nebular attenuation, adopting E(B–V)$_{stars}$ = 0.44 E(B–V)$_{gas}$ from \cite{calz97}. Using this, we compute SFR$_{H\alpha}$ and measure the difference E(B–V)$_{gas}$ – E(B–V)$_{stars}$. We then update E(B–V)$_{gas}$, apply the corresponding dust correction, and recalculate SFR$_{H\alpha}$. This process is repeated until the maps converge. While the Balmer decrement would provide a more direct estimate of E(B–V)$_{gas}$, our redshift range precludes access to the \hb\ line needed for such measurements. Although adopting a constant ratio of E(B–V)$_{stars}$ = 0.44 E(B–V)$_{gas}$ would yield comparable results, our method enables this ratio to vary spatially, better capturing local differences in dust attenuation across the galaxy.}

\begin{figure*}[ht!]
\plotone{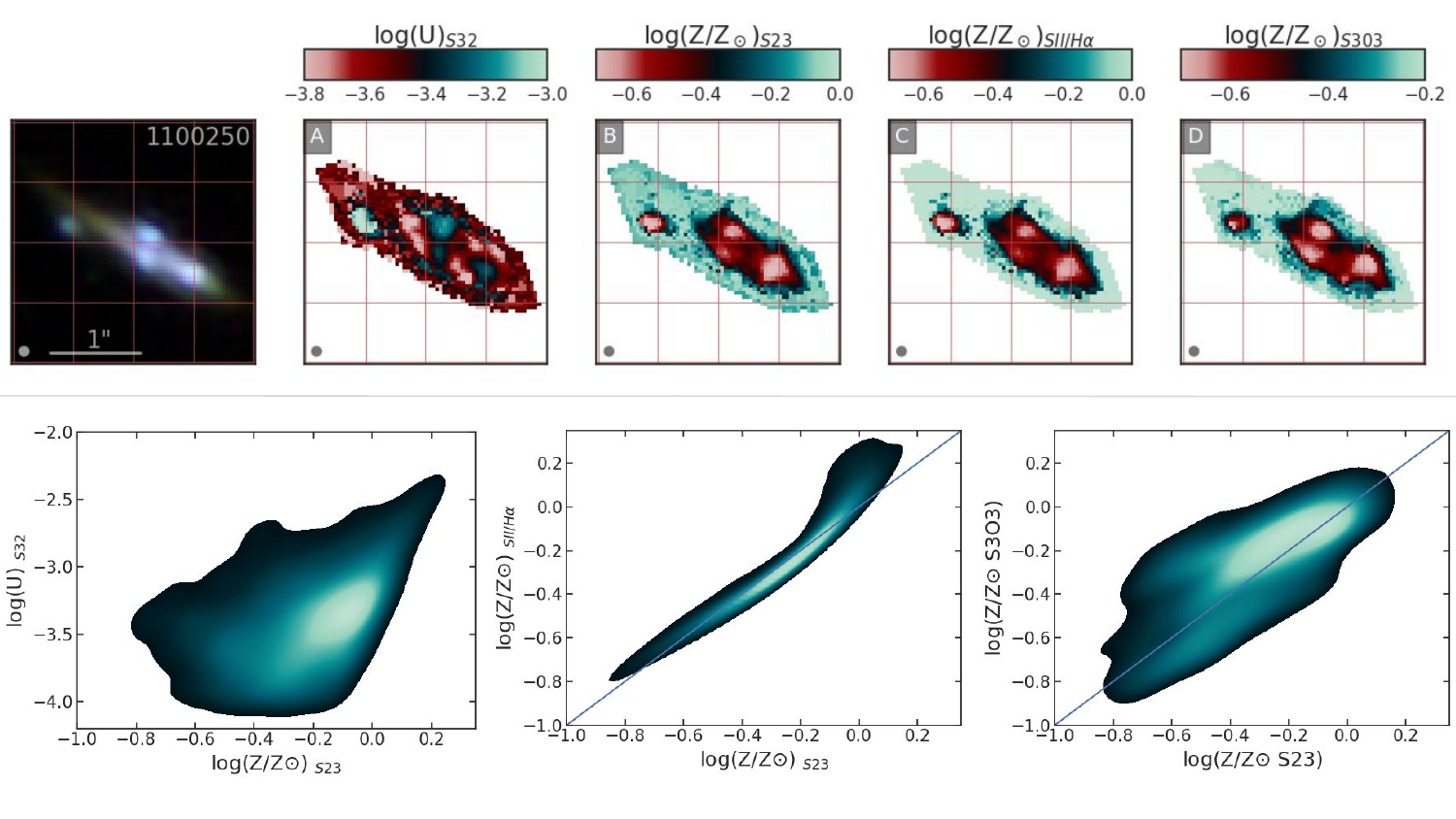}
\caption{ First row - Ionization (log(U) - Panel A) and metallicity maps (\stt, $[SII]/H\alpha$, and \stot\ - Panels B, C, and D respectively) for one of our galaxies (1100250), the first panel also contains a 1" scale and the size of the PSF FWHM. The leftmost panel shows the color image of the galaxy, where we see the four star-forming clumps stand out as bright blue clumps. Panel A shows that regions of clumpy star formation within the galaxy have high ionization, while Panels B, C, and D show that these clumps have a lower metallicity than their surroundings. Note that Panels B and C have been corrected for ionization, while Panel D has not.\editone{Second row - A comparison of the metallicity indicators, all distributions shown here include data for every pixel in every galaxy map. The first panel shows the relationship between the S32 ionization and S23 metallicity. The second panel compares the SII/\ha\ and S23 metallicities, finding a tight correlation. The last panel compares the S3O3 and S23 metallicities, finding a correlation between the indicators. }
\label{fig:metal}}
\end{figure*}

\subsubsection{Masking Non-Star Formation Emission}
\editone{For the final adjustment, we identify regions in the galaxy where emission is not associated with star formation. To do this, we utilize a simplified version of a WHaN diagram \citep{cidf10} (rightmost inset panel of Figure \ref{fig:corr}). This diagram used the equivalent width of \ha\ and log([NII]/\ha) to determine the source of the emission; however, our focus is solely on emissions that originate from star formation. Consequently, we mask regions with \ha\ equivalent widths less than 14 \AA\ because these regions are likely to be contaminated by emission from diffuse ionized gas \citep{lace18,tacc22}. Additionally, we exclude areas where log([NII]/\ha) is greater than -0.4, as these emissions may arise from shocks or active galactic nuclei (AGN) \citep{cidf10}. The regions where the emission is attributed to star formation are shown in the rightmost panel of Figure \ref{fig:corr} for galaxy 4101288.}

\editone{Note that the majority of our galaxies are relatively dust-free and do not require much masking, but here, to illustrate the method,  we show an example of a galaxy that does.}

\subsection{Clump Identification \label{sec:clump}}
To identify clumps in our sample, we use a method similar to techniques used in \cite{guo15, satt23}. We first run a Gaussian smoothing algorithm on both the \ha\ emission line map and the F435W images to model smooth morphological features. We then subtract the smooth model from the data, leaving behind the clumpy structure of the galaxy. We then run source extractor \citep{bert96, barb16} on the residual images to identify the clumpy regions. We use both \ha\ and UV maps as the star-forming clumps will stand out in the \ha\ emission line maps, but lower flux regions can be difficult to identify due to the lower SNR of the \ha\ map, while the UV maps can identify these regions more easily. Additionally, dusty clumps may only be visible in \ha. 

We then combine both segmentation maps (\ha\ and UV).  Once the clumpy features have been segmented, we remove any region that represents the core of the galaxy using the mass maps of the galaxies. We also remove "patchy" regions consisting of noisy flux by setting an SNR limit of 5 in the region.

\subsection{Sleuth}
To extract an emission line map, we must model the galaxy's spatially resolved grism spectra. Current standards for modeling grism spectra assume the galaxy has the same stellar population everywhere, something we know not to be fundamentally true of galaxies. This assumption will lead to over- or under-subtraction of the continuum and nearby emission lines \citep{sorba17,estr24}. This also leads to residual flux that will contaminate the emission line map and affect science downstream. 

To overcome this issue, we have developed our spatially resolved grism modeling software, \Sleuth. An early version of this code is featured in \cite{estr24}. The spatially resolved fitting methods of \Sleuth\ account for spatially varying stellar populations within a galaxy and produce high-quality emission line maps. \Sleuth\ works by first segmenting the galaxy into small regions. \editone{We begin by identifying the brightest pixel and then use a nearest-neighbors approach to group pixels with similar colors. Pixels are added to a region until a target signal-to-noise ratio (SNR = 50 in F150W) is reached. Due to this SNR threshold, many of the resulting regions consist of just a single pixel. Once the galaxy is segmented, we forward-model a set of spectral templates—including emission lines—for each region, and then optimally fit these templates to the observed grism spectra.} This approach allows for the modeling of spatially varying stellar populations and emission line strengths, reducing contamination from continuum flux in emission line maps.

Figure \ref{fig:data} shows an example of the \jwst/NIRISS grism data along with emission line maps extracted using \Sleuth. The top panel shows the GR150C and GR150R dispersions (first and second rows, respectively) for one of the galaxies in our sample (1100250). Note that the emission lines are tilted along the dispersion direction because their morphology reflects the morphology of the star-forming regions that emit them. This galaxy has several spatially resolved emission lines present (indicated with arrows), whose maps are shown in the bottom two rows. Note that the emission line maps shown here have been binned using \sleuth's segmentation feature, which sets the minimum SNR of each region to 10 in the \ha\ emission line map.  Furthermore, several of these lines overlap ([OIII]$\lambda4959$ and [OIII]$\lambda5007$ as well as H$\alpha$ and [SII]), but \Sleuth's spatially resolved modeling returns minimally contaminated emission line maps. Note that our emission line maps have been PSF matched to the F444W resolution (similar to the imaging data used). 

\subsection{Spatially Resolved Metallicity}

For this work, we make use of several emission line maps including  $H\alpha$, $[SII]$, $[SIII]\lambda9069$, $[SIII]\lambda9532$. From this combination of emission lines, we derive metallicity maps utilizing
\begin{equation}
    S23 \equiv \frac{[SIII]\lambda \lambda9069,9532 + [SII]\lambda \lambda6717,6731}{H\alpha},
\end{equation}
with ionization maps generated using 
\begin{equation}
    S32 \equiv \frac{[SIII]\lambda \lambda9069,9532}{[SII]\lambda \lambda6717,6731}.
\end{equation}

These relationships are converted to metallicity/ionization using equations from \cite{kewl19}. In our redshift range, additional metallicity indicators are also available, namely

\begin{equation}
    [SII]/H\alpha \equiv \frac{[SII]\lambda \lambda6717,6731}{H\alpha}
\end{equation}
and
\begin{equation}
    S3O3 \equiv \frac{[SIII]\lambda \lambda9069,9532}{[OIII]\lambda4959,5007}
\end{equation}
using equations from \cite{kewl19} and \cite{stas06} respectively. 

Our preferred metallicity indicator for this work is \stt\ because it is available for a larger number of galaxies (due to its wider redshift coverage) and because it is not highly dependent on ionization, whereas $[SII]/H\alpha$ is \citep{kewl19}. \stt\ is also preferred because it does not suffer from the limitation of lower SNR for [SII].  For these reasons, in Sections 4 and 5 we use \stt\ to analyse our galaxies. However, in this Section, we make comparisons between \stt\ and the other two indicators, where available, to test the robustness of our results.   
Figure \ref{fig:metal} shows an example of the ionization and metallicity maps for the galaxy shown in Figure \ref{fig:data}. Panel A shows the ionization (log(U)) map calculated using \strt. Note that there is a metallicity dependency on ionization, though the \strt\ ratio itself is not very dependent on metallicity \citep{kewl19}, making it an ideal ionization ratio. For the initial metallicity in our fitting algorithm, we use \stot\ (Panel D) when available, and otherwise, when not, we use a guess of log(Z/Z$_\odot$) = 0. \editone{With the log(U) maps in hand, we calculate the \stt\ and [SII]/\ha\ metallicities (Panels B and C, respectively). We then iterate between ionization and metallicity calculations until the metallicity converges. Since \strt\ is only weakly sensitive to metallicity, the choice of initial metallicity has a negligible effect—as long as it is within a realistic range. Different initial guesses yield differences that are well below the typical metallicity uncertainties.}

As can be seen in the top row of Figure \ref{fig:metal} Panels B, C, and D, all metallicity maps for this galaxy look similar, adding confidence to our spatially resolved metallicities. \editone{The second row of Figure \ref{fig:metal} compares \stt\ metallicities to those derived from the other indicators across our full sample. The left panel shows \strt\ ionization versus \stt\ metallicity, revealing a clear correlation. While this trend runs counter to predictions from photoionization models, it is consistent with findings from \citet{dopi14, poet18, ming20, jix22}, where the origin of this discrepancy remains unresolved. The middle panel compares metallicities derived from [SII]/\ha\ and \stt, showing a tight correlation with a median difference of 0.01$^{+0.02}{-0.07}$. The rightmost panel compares S3O3-based and \stt\ metallicities, which also broadly agree but exhibit greater scatter. The median offset in this case is –0.04$^{+0.14}{-0.15}$. Overall, these comparisons suggest that the \stt\ indicator produces metallicities consistent with other strong-line indicators, while also being weakly dependent on ionization. } 

\begin{figure*}[ht!]
\epsscale{1.1}
\plotone{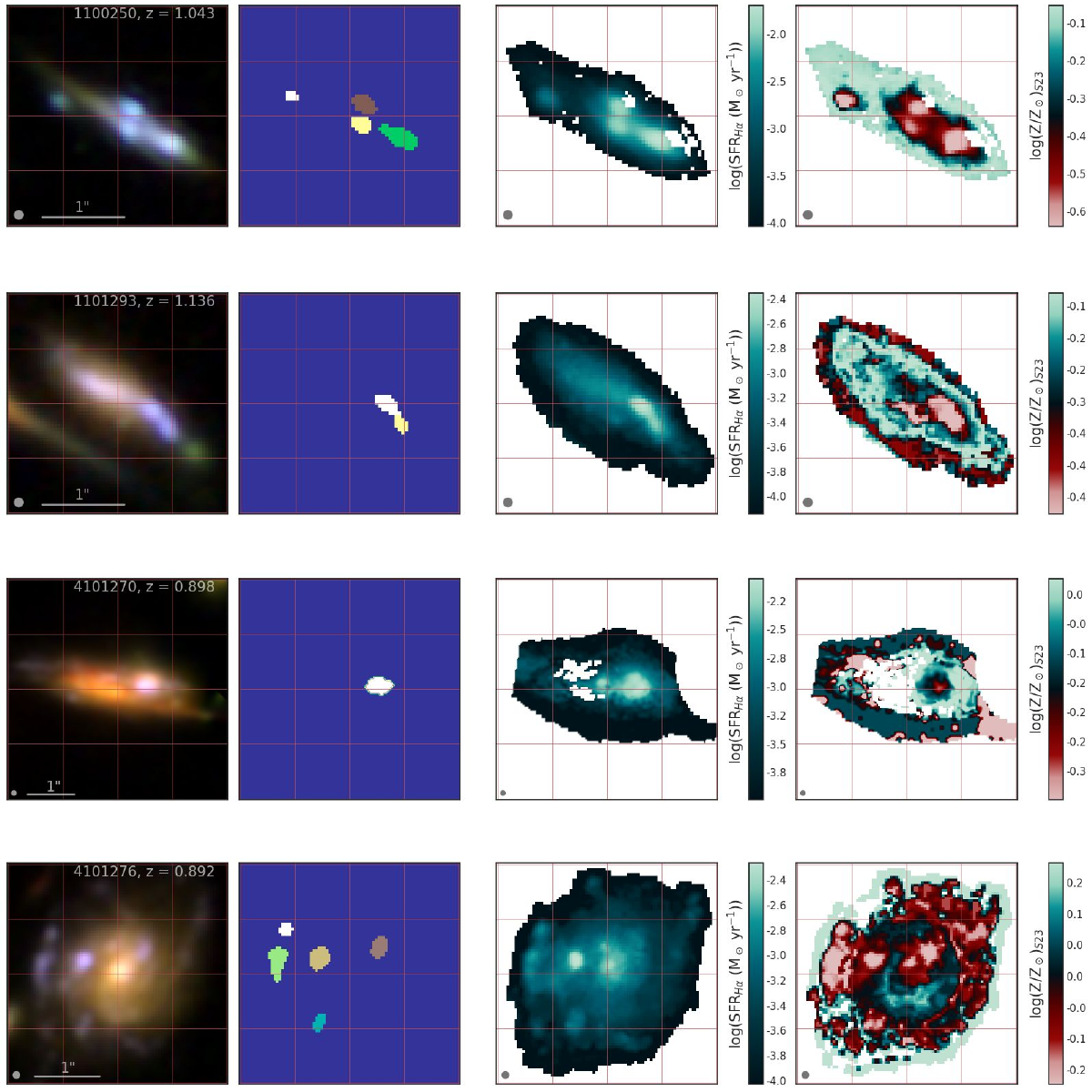}
\caption{Spatially resolved property maps of four example galaxies (all galaxies shown in the Appendix). Each row shows a color image of the galaxy, clump segmentation, log(SFR) map (corrected for lensing), and log(Z/Z$_\odot$) map. The log(SFR) and log(Z/Z$_\odot$) maps were derived using emission line maps. In the first panel we also include a 1" scale and the PSF FWHM}
\label{fig:maps}
\end{figure*}

Table \ref{table1} shows properties for our galaxy sample, including their CANUCS ID, z$_{grism}$, $\mu$ (magnification), \ha\ SFR, stellar mass, sSFR, and \stt\ metallicity. Note that while our galaxies are lensed, emission line ratios are unaffected by lensing. The magnification is shown with the 68-percentile scatter showing that for most of our galaxies, magnification shows little variance as we purposely avoided lensed arcs to avoid highly distorted objects. 

\begin{deluxetable*}{lcccccc}
\tablecaption{Properties of Sample Galaxies\label{table1}}
\tablecolumns{7}
\tablehead{
\colhead{ID}  &
\colhead{z$_{grism}$}  &
\colhead{$\mu$} &
\colhead{log(SFR)} &
\colhead{log(M/M$_\odot$)} &
\colhead{log(sSFR)} &
\colhead{log(Z/Z$_\odot$)}  \\
\colhead{(1)} & 
\colhead{(2)} & 
\colhead{(3)} &
\colhead{(4)} &
\colhead{(5)} &
\colhead{(6)} &
\colhead{(7)} }
\startdata
1101313 & $0.869_{0.002}^{0.002}$ & $6.074_{1.245}^{3.297}$ & $0.990 \pm 0.023$ & $10.717 \pm 0.200$ & $-9.7 \pm 0.2$ & $-0.056 \pm 0.01$ \\
1100250 & $1.043_{0.001}^{0.001}$ & $7.511_{0.196}^{0.217}$ & $0.533 \pm 0.010$ & $9.013 \pm 0.832$ & $-8.5 \pm 0.8$ & $-0.333 \pm 0.03$ \\
1100417 & $1.043_{0.008}^{0.005}$ & $2.412_{0.020}^{0.029}$ & $-0.371 \pm 0.111$ & $9.926 \pm 0.257$ & $-10.3 \pm 0.3$ & $0.032 \pm 0.03$ \\
1100401 & $1.319_{0.002}^{0.001}$ & $2.582_{0.057}^{0.098}$ & $0.500 \pm 0.062$ & $10.319 \pm 0.364$ & $-9.8 \pm 0.4$ & $-0.322 \pm 0.07$ \\
1100429 & $1.045_{0.006}^{0.007}$ & $3.667_{0.102}^{0.125}$ & $-0.257 \pm 0.085$ & $10.031 \pm 0.249$ & $-10.3 \pm 0.3$ & $-0.170 \pm 0.09$ \\
1100154 & $0.868_{0.015}^{0.015}$ & $1.349_{0.005}^{0.006}$ & $0.216 \pm 0.079$ & $10.521 \pm 0.166$ & $-10.3 \pm 0.2$ & $0.050 \pm 0.02$ \\
1101293 & $1.138_{0.003}^{0.002}$ & $3.214_{0.074}^{0.083}$ & $-0.109 \pm 0.061$ & $9.687 \pm 0.335$ & $-9.8 \pm 0.3$ & $-0.130 \pm 0.03$ \\
1101318 & $1.100_{0.006}^{0.002}$ & $1.565_{0.006}^{0.008}$ & $0.141 \pm 0.073$ & $10.060 \pm 0.325$ & $-9.9 \pm 0.3$ & $-0.059 \pm 0.04$ \\
5105148 & $1.129_{0.001}^{0.004}$ & $1.251_{0.004}^{0.004}$ & $-0.153 \pm 0.072$ & $9.416 \pm 0.392$ & $-9.6 \pm 0.4$ & $-0.105 \pm 0.05$ \\
5115336 & $1.021_{0.002}^{0.004}$ & $1.757_{0.057}^{0.092}$ & $1.214 \pm 0.021$ & $11.018 \pm 0.205$ & $-9.8 \pm 0.2$ & $-0.193 \pm 0.01$ \\
5114355 & $1.093_{0.028}^{0.023}$ & $1.928_{0.011}^{0.011}$ & $-0.894 \pm 0.094$ & $8.605 \pm 0.375$ & $-9.5 \pm 0.4$ & $-0.211 \pm 0.013$ \\
4101270 & $0.886_{0.001}^{0.002}$ & $2.272_{0.063}^{0.134}$ & $1.007 \pm 0.013$ & $10.150 \pm 0.299$ & $-9.1 \pm 0.3$ & $-0.052 \pm 0.01$ \\
4101276 & $0.892_{0.007}^{0.006}$ & $1.518_{0.011}^{0.012}$ & $0.632 \pm 0.038$ & $10.255 \pm 0.245$ & $-9.6 \pm 0.2$ & $-0.083 \pm 0.01$ \\
4101288 & $0.752_{0.003}^{0.002}$ & $1.561_{0.021}^{0.024}$ & $1.032 \pm 0.029$ & $11.261 \pm 0.125$ & $-10.2 \pm 0.1$ & $-0.113 \pm 0.01$ \\
4101303 & $1.318_{0.001}^{0.004}$ & $3.325_{0.112}^{0.125}$ & $0.795 \pm 0.028$ & $9.737 \pm 1.219$ & $-8.9 \pm 1.2$ & $0.016 \pm 0.01$ \\
4101323 & $1.328_{0.006}^{0.001}$ & $1.303_{0.007}^{0.006}$ & $0.745 \pm 0.036$ & $11.139 \pm 0.129$ & $-10.4 \pm 0.1$ & $0.015 \pm 0.01$ \\
4100141 & $0.642_{0.002}^{0.007}$ & $1.119_{0.002}^{0.002}$ & $0.906 \pm 0.036$ & $10.461 \pm 0.201$ & $-9.6 \pm 0.2$ & $-0.320 \pm 0.03$ \\
4100238 & $1.210_{0.005}^{0.001}$ & $2.111_{0.034}^{0.037}$ & $0.516 \pm 0.041$ & $10.539 \pm 0.190$ & $-10.0 \pm 0.2$ & $-0.219 \pm 0.03$ \\
4100575 & $1.082_{0.002}^{0.002}$ & $2.119_{0.021}^{0.028}$ & $-0.111 \pm 0.033$ & $8.941 \pm 0.440$ & $-9.1 \pm 0.4$ & $-0.232 \pm 0.02$ \\
4100186 & $1.316_{0.002}^{0.001}$ & $1.603_{0.005}^{0.005}$ & $-0.283 \pm 0.072$ & $9.383 \pm 0.553$ & $-9.7 \pm 0.6$ & $-0.372 \pm 0.027$ \\
\enddata
\tablecomments{(1) catalog ID number (matching those in Sarrouh, Asada et al. in prep); (2) grism-based redshift;(3) mean magnification, averaged over the galaxy and 68-percentile scatter, derived from the best-fit strong lensing model; (4) total log(SFR) derived from \ha\ maps; (5) total stellar mass from spatially resolved fits to the broadband data; (6) log(sSFR) from \ha\ based SFR and broadband based stellar mass; (7) \stt\ based gas-phase metallicity }
\end{deluxetable*}

\section{Results \label{sec:res}}

\subsection{Metallicity of Star-Forming Clumps}

In Figure \ref{fig:maps}, we present several galaxies from our sample. Each row features a color image of the galaxy in the first panel, followed by a segmentation map of the star-forming clumps described in Section \ref{sec:clump} in the second panel. The third panel displays the \ha\ star formation rate (SFR) maps, and the fourth panel shows the \stt\ metallicity maps. Additional galaxies are included in the Appendix.

The color images here were created by linearly interpolating between broadband images at rest wavelengths of 2800 \AA\ (blue), 6564 \AA\ -  (green), and 18500 \AA\ - (red). These wavelengths were chosen as they represent roughly the bluest common rest wavelength, which lies in the rest UV and correlates to star formation on 100 Myr timescales, the \ha\ wavelength, and the reddest common rest wavelength, which is dominated by low-mass stars that make up most of the stellar mass budget.  


The top 3 rows in Figure \ref{fig:maps} show galaxies with distinct star-forming clumps; four in the top row, two in the outskirts of the second row galaxy, and one in the galaxy in the third row. In these simple systems, it is clearly evident that these star-forming clumps have lower metallicities. The last row illustrates a face-on spiral galaxy with several star-forming clumps. These clumps are prominent in the SFR maps as areas of high star formation, and they display lower metallicities than the surrounding disk in the corresponding metallicity maps. Together, these maps allow us to study Cosmic Noon galaxy properties at high spatial resolution.

\begin{figure*}[ht!]
\epsscale{1.15}
\plotone{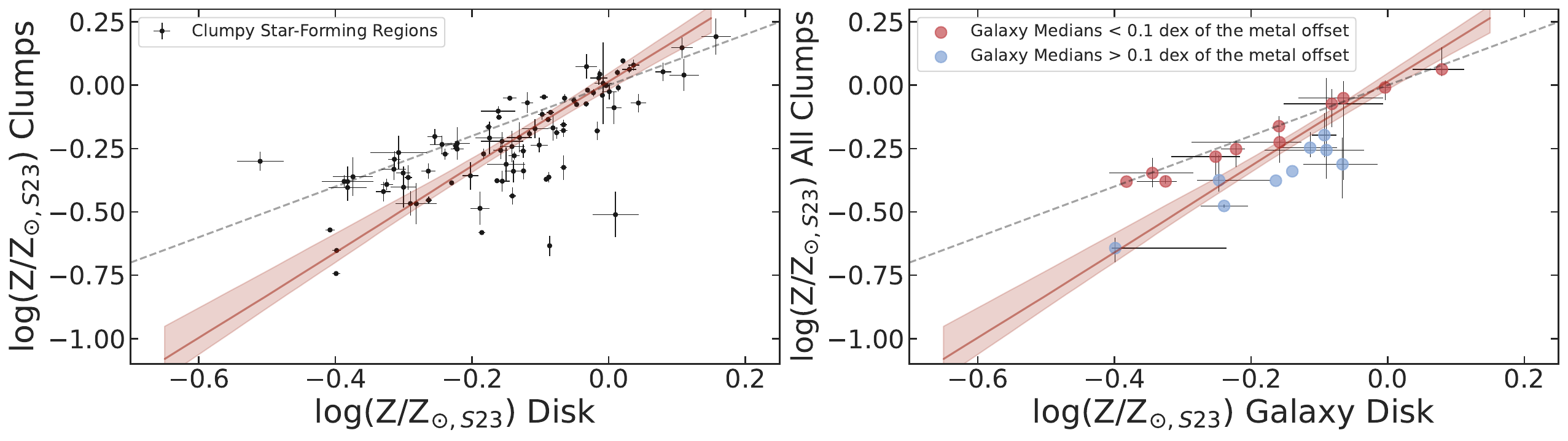}
\caption{A comparison between the metallicities of the star-forming clumps to the region of the disk surrounding the clump. In the left panel, we show the metallicities for each star-forming clump and the nearby disk, while the right panel shows median metallicity for each galaxy with the error bars representing the scatter in the sample (for galaxies with one star-forming clump, the value of that clump is shown with no error bar). In both panels, we show a linear fit to the individual clumps and a dashed 1-to-1 line that represents clumps that have the same metallicity as the galactic disk that surrounds them.  Here, we see that most of our clumps have a lower metallicity than their surroundings. In the right panel, we have highlighted galaxies whose median values sit closer to the 1-to-1 line ($<$ 0.1 dex metallicity offset, red points). These galaxies may be forming clumps through a different mechanism than galaxies that have a larger offset (blue), though a larger sample would be necessary to determine what the differences are. 
\label{fig:clump_to_disk}} 
\end{figure*}

To compare the metallicities of the star-forming clumps with those of the surrounding galaxy, we first measured the metallicity within the clumps and areas just outside the clumps using a circular aperture of 4 pixels (the median size of our clumps) and an annulus 4 pixels wide separated by 2 pixels from the inner aperture, for each respective region. For the aperture and annulus, we mask out any nearby clumps. The results of these comparisons are shown in Figure \ref{fig:clump_to_disk}. In the left panel, we present the individual clumps' \stt\ metallicities, along with a 1:1 dashed line and a linear fit of the data (solid line) along with the 68th-percentile confidence region (shaded region). The right panel displays the median values for the clumps within each galaxy, with error bars indicating the scatter within the populations. The linear fits indicate that clumps generally have lower metallicities than their surrounding galaxy regions, with a mean offset of $\sim$ 0.1 dex, indicating that gas in the star-forming clump is diluted by $\sim$ 20 $\%$. Note that we see similar results if we compare the clump metallicity to the overall disk (with clumps masked out) metallicity.

In the right-hand panel of Figure \ref{fig:clump_to_disk}, which shows the galaxy median values, there is evidence of two potential galaxy populations. Galaxies whose clumps have similar metallicity to their surrounding disk (< 0.1 dex metallicity offset) shown in red; these galaxies sit near the 1-to-1 line, and galaxies whose clumps have larger offsets (shown in blue). What we are potentially seeing here is that galaxies with a smaller metallicity offset (shown in red) may predominantly form their clumps through disk instabilities and therefore form their clumps from gas present in the galaxy, leading to a smaller or non-existent offset. While the larger offset galaxies (shown in blue) may have their clumps form primarily from metal-poor gas falling into the galaxy, fueling star formation and diluting metallicity. We note that clump populations within the individual galaxies do show a large scatter of metallicity offset values, and therefore, we need a larger sample of galaxies in order to make any further determinations.

\section{Discussion \label{sec:dis}}
\subsection{The Link Between SFR and Metallicity}

\begin{figure*}[ht!]
\epsscale{1.1}
\plotone{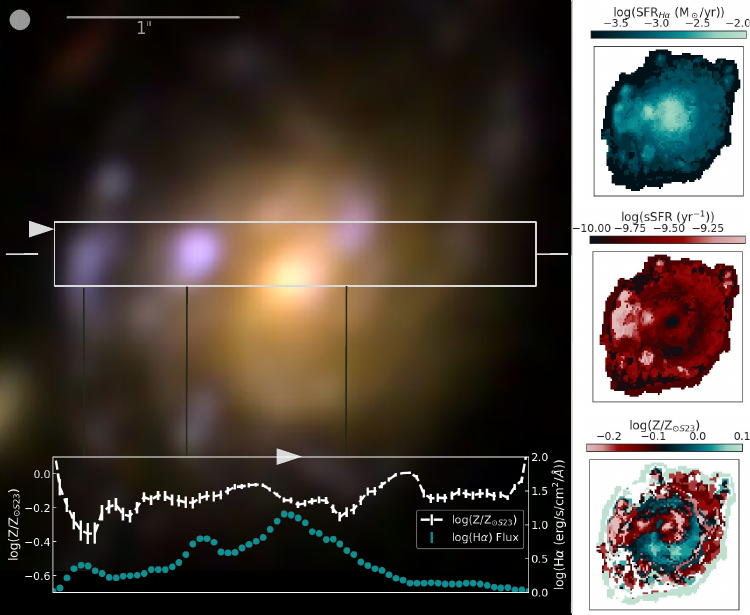}
\caption{ Example galaxy (4101276) showing the distribution of metallicity and \ha. The galaxy is shown in the top left panel, where the star-forming clumps are seen as small blue regions over the redder galaxy. The right panels show the spatially resolved SFR$_{\ha}$, sSFR (derived from \ha), and metallicity. The bottom panel shows the \ha\ and metallicity of the strip outlined on the galaxy. We can see from the maps and distribution of metallicity and \ha\ that these two properties are anti-correlated, especially in the star-forming clumps. 
\label{fig:direct_compare}}
\end{figure*}

\begin{figure*}[ht!]
\epsscale{1.15}
\plotone{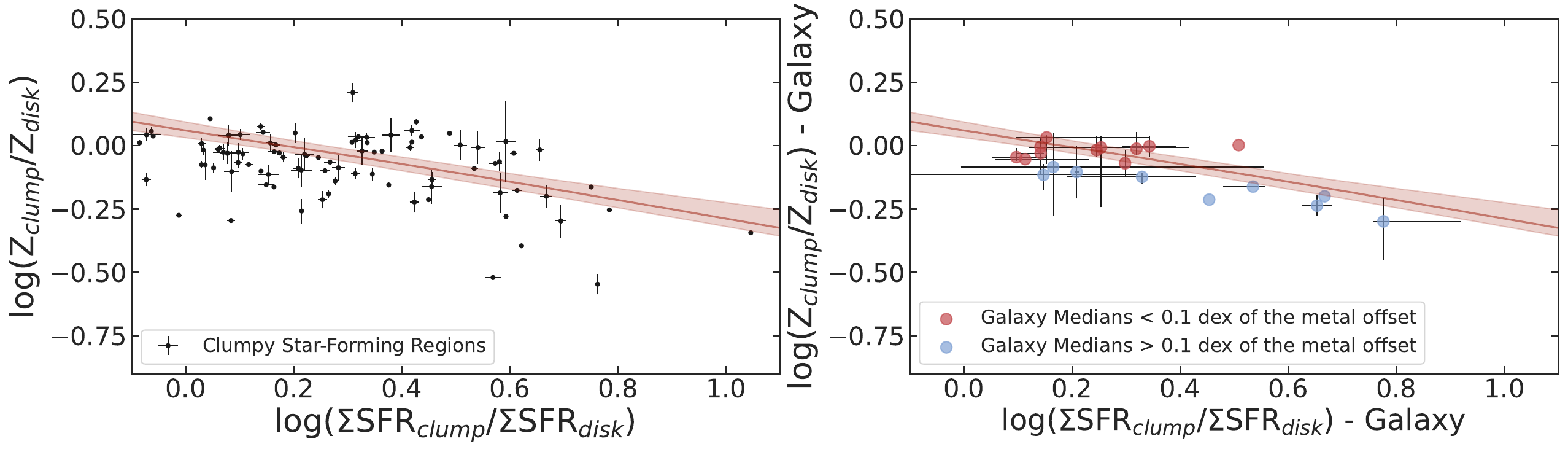}
\caption{A comparison of metallicity to SFR in the star-forming clumps for each of our metallicity indicators.  In the left panel, we show values for each star-forming clump, while the right panel shows median values for each galaxy with the error bars representing the scatter in the sample. In both panels, we show a linear fit to the individual clumps, which shows that the more elevated the clump SFR is compared to the disk, the lower the clump's metallicity is compared to the surrounding disk.
\label{fig:sfr_to_metal}}
\end{figure*}

Our SFR and metallicity maps (see Figure \ref{fig:maps} and the Appendix) reveal that the dilution of metallicity in star-forming clumps is linked to an increase in star formation rate. To explore this connection further, we examine a specific galaxy illustrated in Figure \ref{fig:direct_compare}. In the color image of the galaxy, several star-forming clumps appear in blue against a face-on, comparatively red galaxy. 

To assess the correlation between SFR and metallicity in Figure \ref{fig:direct_compare}, we select a strip of the galaxy, summing its \ha\ flux and measuring the metallicity along the strip. This strip includes three star-forming clumps as well as a portion of the galaxy's bulge. The bottom left panel of Figure \ref{fig:direct_compare} directly compares metallicity to \ha\ flux (and thus SFR), the panel is lined up with the color image to show where the flux/metallicity measurement is coming from. 

From this comparison, we observe that for the leftmost two clumps, the peaks in \ha\ flux coincide with the local minima in metallicity, demonstrating an anti-correlation between metallicity and SFR. Additionally, this indicates that these clumps contain gradients in both SFR and metallicity. Not only does each clump exhibit a higher SFR and lower metallicity, but the region with the highest SFR density also corresponds to the lowest metallicity. 

Towards the center of the galaxy, we find the bulge adjacent to a nearby star-forming clump. The \ha\ flux distribution shows that the bulge emits a large amount of \ha\ flux, overpowering the nearby clump's \ha\ flux. However, the metallicity does not reach its local minima where the \ha\ flux of the bulge peaks; instead, the local minima correlate with the location of the star-forming clump. This indicates that while there is an anti-correlation between SFR and metallicity, the mechanism driving star formation in the bulge differs from that of the star-forming clumps.

In Figure \ref{fig:direct_compare}, the three panels on the right display SFR$_{\ha}$, specific star formation rate (sSFR), and metallicity maps. A comparison of these three maps reveals that the star-forming clumps exhibit an elevated SFR and sSFR alongside decreased metallicity. In contrast, the bulge presents higher SFR, lower sSFR, and slightly decreased metallicity than the disk of the galaxy. Consequently, while the bulge does show elevated \ha\ emission, indicating a higher star formation rate, it is comparatively less active in terms of star formation than the star-forming clumps. This disparity further reinforces that the mechanism governing the star formation in the bulge is distinct from what drives the star-forming clumps.

Figure \ref{fig:sfr_to_metal} illustrates the comparison of $\Delta$ log(Z)(log$\frac{Z_{clump}}{Z_{disk}}$) with $\Delta$ star-formation rate density (log$\frac{\Sigma SFR_{clump}}{\Sigma SFR_{disk}}$). The individual clumps are displayed in the left-hand panel, while the medians per galaxy are presented in the right-hand panel, with population scatter indicated by error bars and colors referring to the metallicity offset as seen in Figure \ref{fig:clump_to_disk}. We observe an anti-correlation between $\Delta$ metallicity and $\Delta$ $\Sigma$SFR: clumps with a larger offset in $\Sigma$SFR tend to have a larger offset in metallicity. Therefore, we conclude that the mechanism decreasing the metallicities in the star-forming clumps is also causing these regions to exhibit higher SFR densities.

\begin{figure*}[ht!]
\plotone{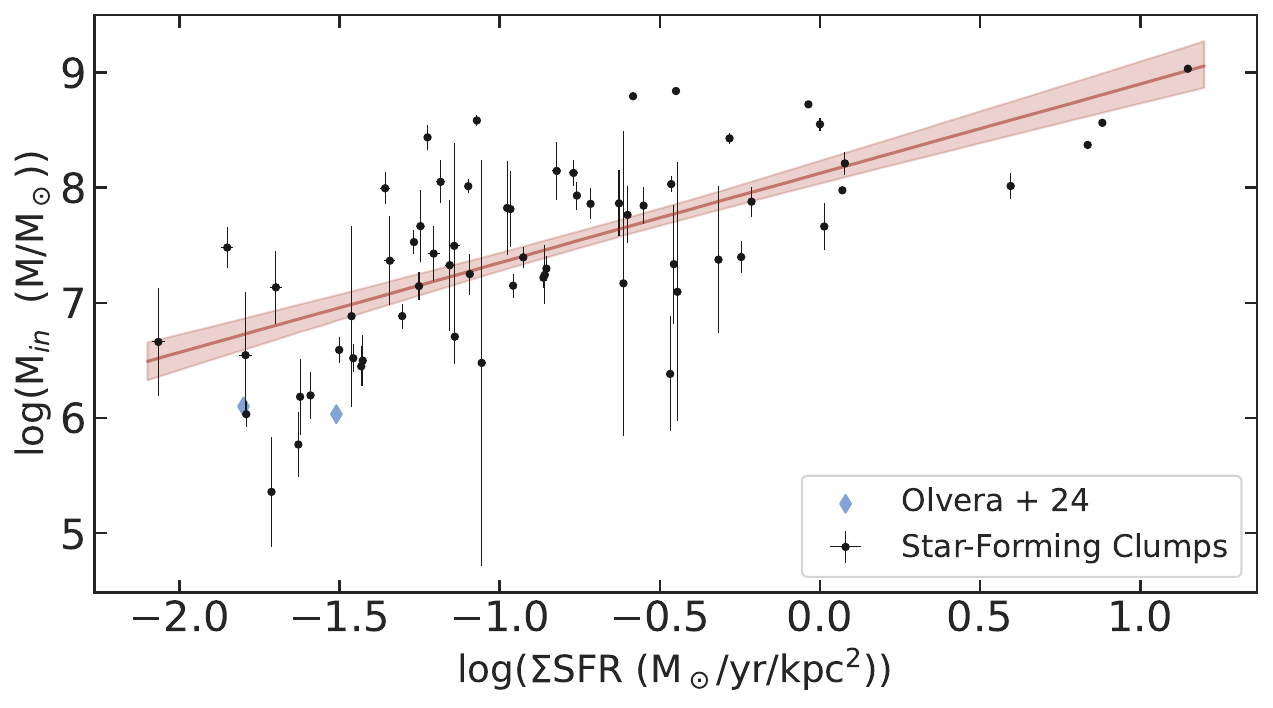}
\caption{Here we show the mass of the infalling metal-poor gas needed to cause the dilution in metallicity we see for each star-forming clump using the simple model from \cite{mann10}, along with a linear fit to the clumps. Using equation \ref{eq:infall}, we estimate the amount of metal-poor gas needed to recreate the metallicity offsets experienced by the star-forming clump vs the star formation in the clump. We can see that regions experiencing high levels of star formation require more metal-poor gas. We also include measurements from NGC99 from \cite{olve24}.
\label{fig:tracks}}
\end{figure*}

We believe that the changes in SFR and metallicity are driven by inflowing metal-poor gas. This inflow of metal-poor gas would increase the SFR while simultaneously lowering the gas-phase metallicity. Using the simple infall model from \cite{mann10},
\begin{equation}
    Z_{obs} = Z_{init} - log\left( 1 + \frac{M_{in}}{f_g M_\odot} \right),
    \label{eq:infall}
\end{equation}
where the dilution of metallicity is solely caused by metal-poor gas, we can estimate the mass of the infalling metal-poor gas. In equation \ref{eq:infall}, $Z_{obs}$ is the observed clump metallicity, $Z_{init}$ is the pre-diluation metallicity (here we assume the metallicity of the nearby disk), $M_{in}$ is the mass of infalling metal-poor gas, and $f_g$ is the gas fraction, set to 0.3 (following \cite{mann10}). Figure \ref{fig:tracks} shows the relationship between infalling gas and $\Sigma$SFR, with individual clumps shown alongside a linear fit to the data. We observe a clear correlation between the SFR density of the clump and the mass of the infalling metal-poor gas. We also include measurements from NGC99 from \cite{olve24}, where we see that the measurements are consistent with ours. We do not speculate on how these measurements compare to our own, as the methods to estimate gas mass are very different. The results shown in Figure \ref{fig:tracks} are consistent with findings from MANGA at 0.1 $<$ z $<$ 0.17 \citep{barr18}, which indicate that gas-phase metallicity decreases as the gas fraction increases. Consequently, a sudden influx of gas mass would lead to a reduction in metallicity. These results lend confidence to the hypothesis that metal-poor gas acts as fuel for these star-forming clumps, diluting metallicity and increasing SFRs. 

\subsection{Clumps or Companions?}

Several works \citep{mowl22, sues23, mowl24, whit25} have seen the presence of smaller bodies (globular clusters or satellite galaxies) near larger galaxies. These works show galaxies growing through the accretion of smaller bodies. It may be possible that what we are seeing in our work is the accretion of satellite galaxies onto larger galaxies, and 
that these star-forming clumps may not have been formed in the galaxy by disk instabilities \citep{ager09, deke09, ceve10} or by infalling clumps of metal-poor gas, but rather are satellite galaxies seen in projection. Frustratingly, we should see the same "dilution" of metallicity if this were the case, as simply following the mass-metallicity relationship \citep{trem04} indicates that lower mass satellite galaxies should have lower metallicities than the larger mass main galaxy. 

\begin{figure*}[ht!]
\epsscale{1.2}
\plotone{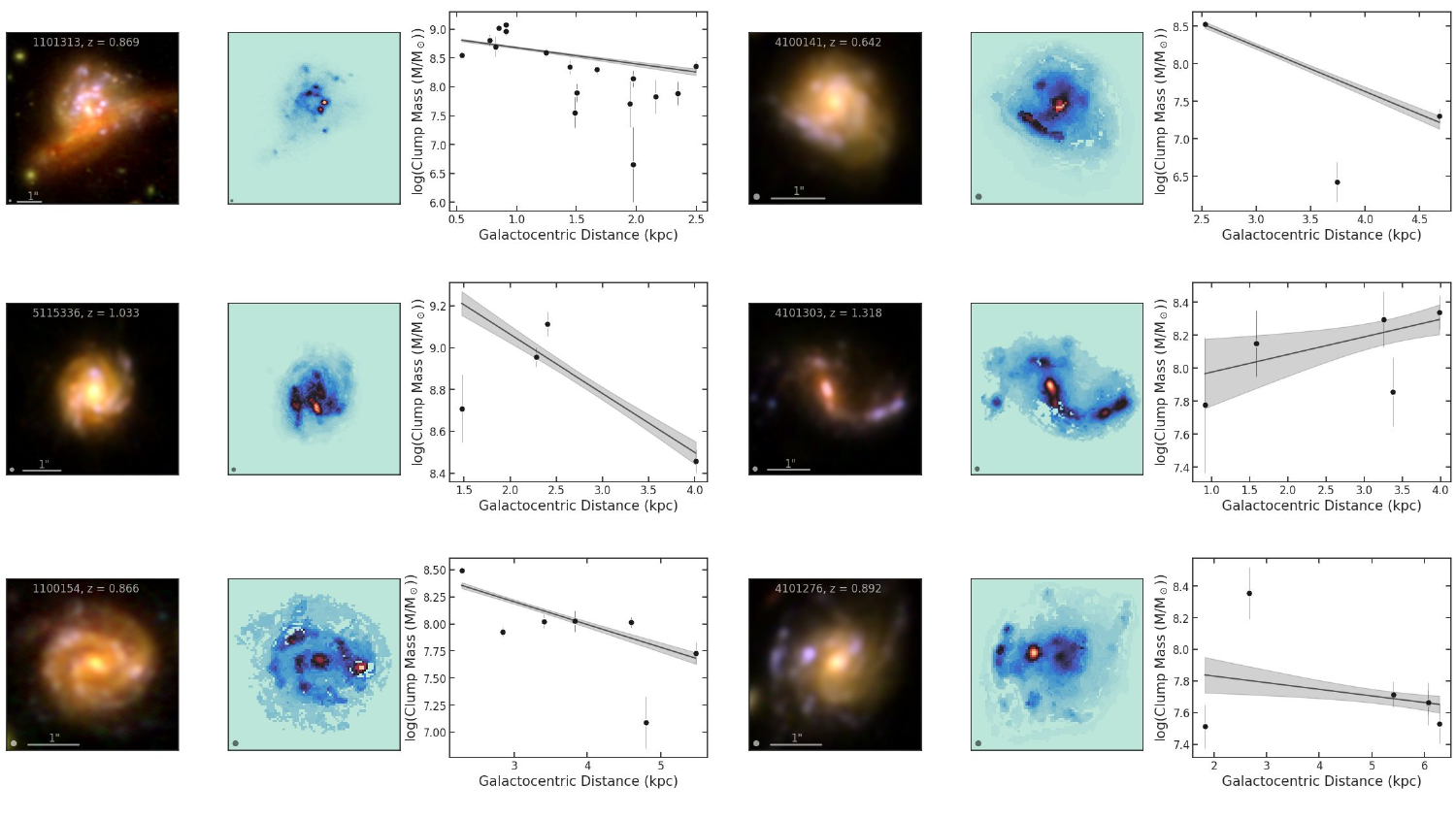}
\caption{In each panel, we show a color image of one of our face-on galaxies, the uncorrected \ha\ map (to better visualize the location of clumps), and the clump mass vs galactocentric distance. We see that, in general, clumps closer to the center of the galaxies have higher masses. This result supports the idea that clumps formed in the galaxy and are migrating towards the center. 
\label{fig:dist}}
\end{figure*}

Figure \ref{fig:dist} attempts to determine if our clumps formed in the galaxy or were accreted satellites. Several works have connected the radial change in clump properties to clump migration towards the bulge of the galaxy  \citep{guo12,shib16,guo18,soto17,huer20}. One common property seen is the increase in clump mass towards the center of the galaxy \citep{kali25}. To test if we see a similar galactocentric dependence in our galaxies (and therefore are seeing clumps formed within the galaxy) we examine six of our face-on galaxies (so as to avoid the complications of inclination) and measure the galactocentric distances of each of the clumps as well as their masses (using aperture photometry to remove the underlying galaxy light). We fit each clump using the same SED fitting process outlined in Section \ref{sec:bbfit}. In Figure \ref{fig:dist} each panel shows a color image of the galaxy, the uncorrected \ha\ map (as to better see the location of the clumps), and the clump mass vs galactocentric distance (along with linear fit). In general, we see that clumps closer to the center of the galaxy have higher masses. These results do not rule out the possibility of accreting satellites, but add some confidence that these clumps formed in situ.

Almost certainly, a few of our star-forming clumps are satellites, as a few stand out in our stellar mass maps. However, the only real way to differentiate between satellites and star-forming clumps is through dynamics, which NIRISS does not have the spectral resolution to extract. \jwst/NIRSpec IFU data may be the only way to disentangle this problem. 

\section{Summary}
This work focuses on the spatially resolved analysis of galaxies with clumpy star formation at 0.6 < z < 1.35 using data from the CAnadian NIRISS Unbiased Cluster Survey (CANUCS). Using slitless grism spectra and broadband imaging, we produced detailed emission line maps, specifically examining \ha, [SII], and [SIII] emission lines. 

By examining \stt\ metallicity maps, we compared the metallicities of star-forming clumps to their local galactic environment (rather than the overall galaxy metallicity, as metallicity gradients may exist). Our findings are as follows:

\begin{itemize}
    \item On average, star-forming clumps have lower metallicities than the surrounding galaxy
    \item The mean offset is $\sim$ 0.1 dex lower, indicating that gas in the star-forming clump has been diluted by $\sim$ 20 $\%$
    \item \editone{We see evidence of two potential formation pathways for our star-forming clumps. (1) They form using internal gas reservoirs, (2) they form from infalling metal-poor gas}
\end{itemize}

We then examined the relationship between SFR and metallicity within the star-forming clumps. We found:

\begin{itemize}
    \item Clumps themselves contain SFR and metallicity gradients, which are negative in the case of SFR and positive for metallicities
    \item We see that the change in SFR is anticorrelated to the change in metallicity within the clumps
\end{itemize}

We then addressed the possibility that the star-forming clumps we examined may have been satellite galaxies seen in projection. We found:

\begin{itemize}
    \item Our face on galaxies showed a trend of higher mass clumps being in closer proximity to the center of the galaxy, which is expected for clumps formed in situ that migrate towards the bulge
    \item To make a more certain statement on the origins of our clumps we would need to turn towards \jwst\ NIRSpec IFU 
\end{itemize}

Our findings indicate that star-forming clumps within these galaxies exhibit lower metallicities than their surroundings, suggesting the inflow of metal-poor gas. Inflowing metal-poor gas would enhance star formation rates while diluting metallicity. This research emphasizes the significance of understanding the complex interplay between star formation and chemical enrichment at Cosmic Noon and provides insights into the mechanisms that govern galaxy formation.

\begin{acknowledgements}
This research was enabled by grants 18JWSTGTO1 and 23JWGO2B04 from the Canadian Space Agency and a Discovery Grant from the Natural Sciences and Engineering Research Council of Canada. MB acknowledges support from the ERC Grant FIRSTLIGHT, the Slovenian national research agency ARRS through grants N1-0238 and P1-0188, and the program HST-GO-16667, provided through a grant from the STScI under NASA contract NAS5-26555. VEC acknowledges support from the Beus Center for Cosmic Foundations. The MAST DOI for CANUCS is \href{https://doi.org/10.17909/ph4n-6n76}{doi:10.17909/ph4n-6n76}. This research used the Canadian Advanced Network For Astronomy Research (CANFAR) operated in partnership with the Canadian Astronomy Data Centre and The Digital Research Alliance of Canada, with support from the National Research Council of Canada, the Canadian Space Agency, CANARIE, and the Canadian Foundation for Innovation. 
\end{acknowledgements}

\appendix 
\begin{figure*}[ht!]
\epsscale{1.0}
\plotone{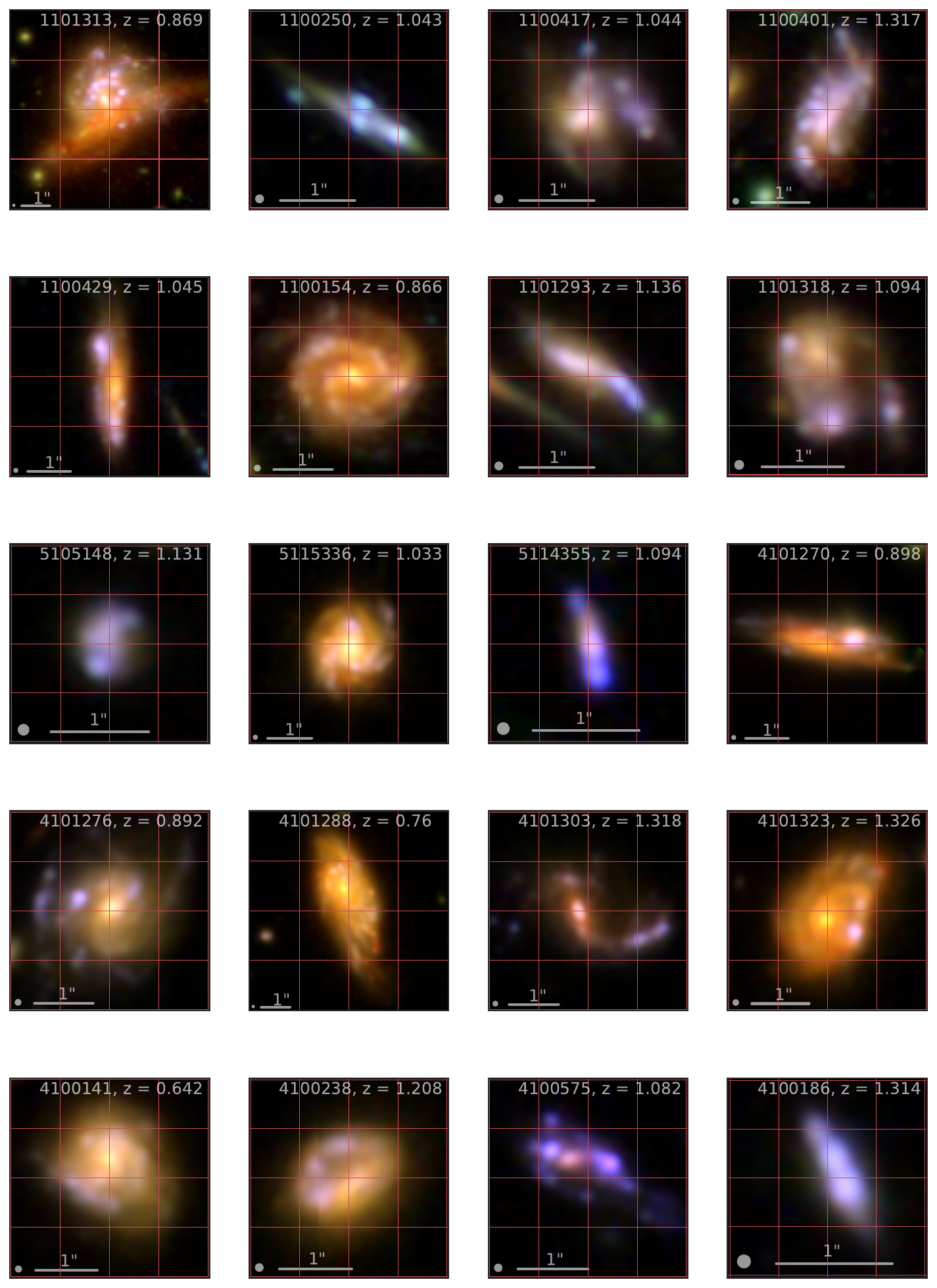}
\caption{Color images of all our galaxies
\label{fig:a_img}}
\end{figure*}

\begin{figure*}[ht!]
\epsscale{1.0}
\plotone{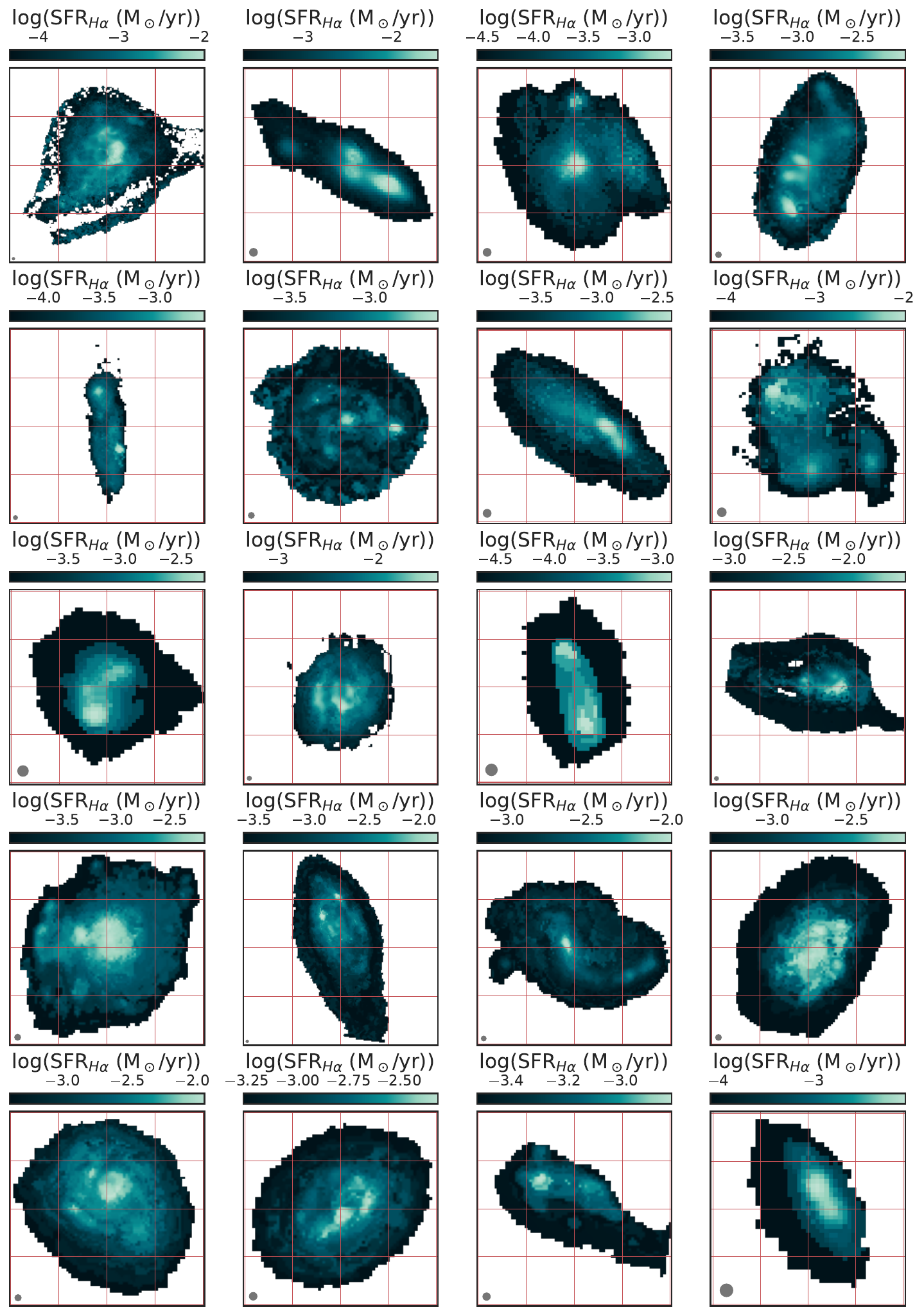}
\caption{\ha\ SFR maps for our entire sample. 
\label{fig:a_sfrs}}
\end{figure*}

\begin{figure*}[ht!]
\epsscale{1.0}
\plotone{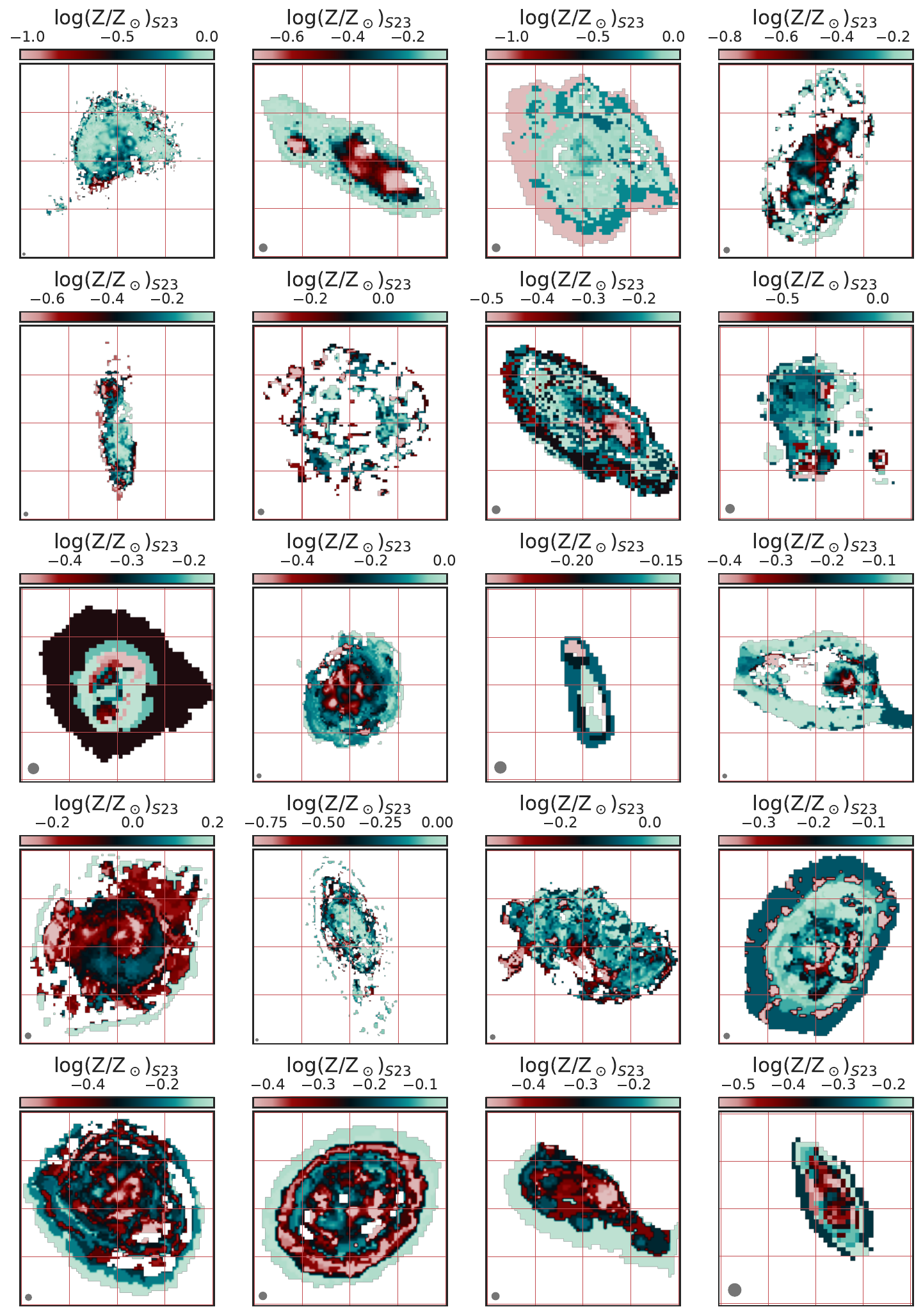}
\caption{All metallicity for our sample.
\label{fig:a_zs}}
\end{figure*}

\bibliography{library}{}
\bibliographystyle{aasjournal}

\end{document}